\documentclass[11pt,twoside]{article}
\usepackage{asp2004}
\usepackage{psfig}
\usepackage{epsf}
\usepackage{graphics}
\usepackage{lscape}
\def\third{{3$^{\rm rd}$}}
\def\deg{{$^{\circ}$}}
\def\bd17{\mbox{BD +17\deg 3248}}
\def\cs22{\mbox{CS 22892-052}}

\def\etal{\mbox{et al.}}

\def\edcomment#1{\iffalse\marginpar{\raggedright\sl#1\/}\else\relax\fi}
\reversemarginpar

\begin{document}
\title{ 
Neutron-Capture Element Abundances in Halo Stars}
\author{John J. Cowan}
\affil{Department of Physics \& Astronomy, University of Oklahoma, Norman, OK
73019}
\author{Christopher Sneden}
\affil{
Department of Astronomy and
McDonald Observatory, University of Texas,
Austin, Texas 78712
}

\begin{abstract}
We present new abundance observations of neutron-capture elements in
Galactic stars. These include new Hubble Space Telescope (HST) detections of
the elements Ge, Zr and Pt in a group of 11 halo stars.
Correlations between these elements and Eu (obtained with ground-based
observations), and with respect to metallicity,  are  also presented.
\end{abstract}
\thispagestyle{plain}

\section{Introduction}

Abundance comparisons of neutron-capture elements -- those  
formed in slow ($s$-process)
and rapid ($r$-process) neutron-capture nucleosynthesis -- 
are providing important new information  about the nature of the earliest 
Galactic nucleosynthesis and the first stellar generations
(see recent reviews by Truran et al. 2002, Sneden \& Cowan 2003 
and Cowan \& Thielemann 
2004).   
Most previous studies have been restricted to elements detectable with
ground-based observations. We have recently completed a long-term study
of the $n$-capture elements Ge, Zr, Os  and Pt -- all of these elements with
dominant atomic transitions in the UV -- in a  sample of 11 metal-poor 
Galactic halo
stars using the Hubble Space Telescope,  
HST (Cowan et al. 2004). These observations were supplemented with 
Keck~I High Resolution Echelle Spectrograph (HIRES) observations of the
$n$-capture element Ir.
Our focus has been to
examine correlations among these elements in an attempt to help
to identify the sites of nucleosynthesis for these elements and also to
trace their evolution with Galactic metallicity.

\section{Neutron-Capture Element Abundance Comparisons}

\begin{figure}[!ht]
\epsfxsize=5.25in
\plottwo{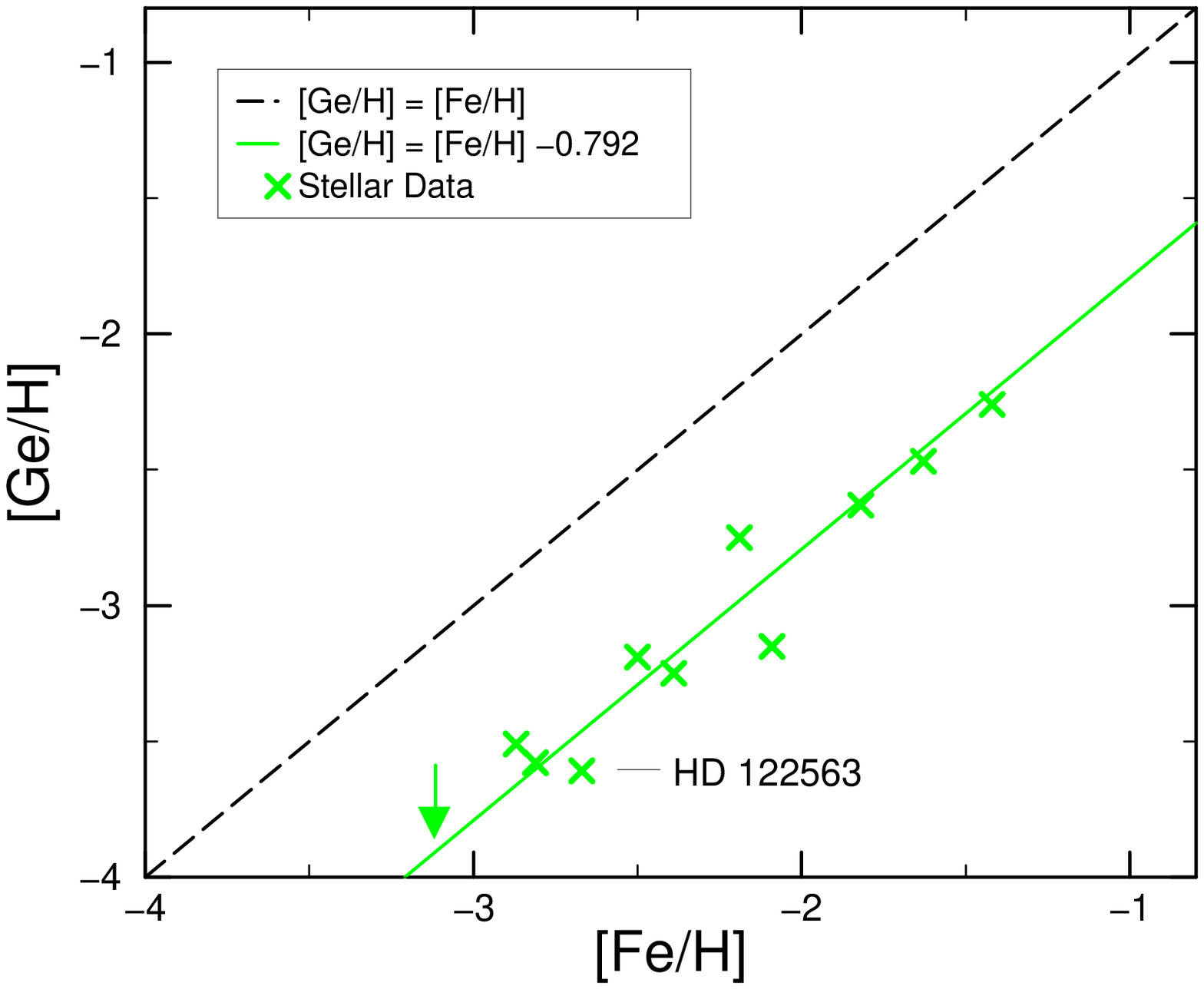}{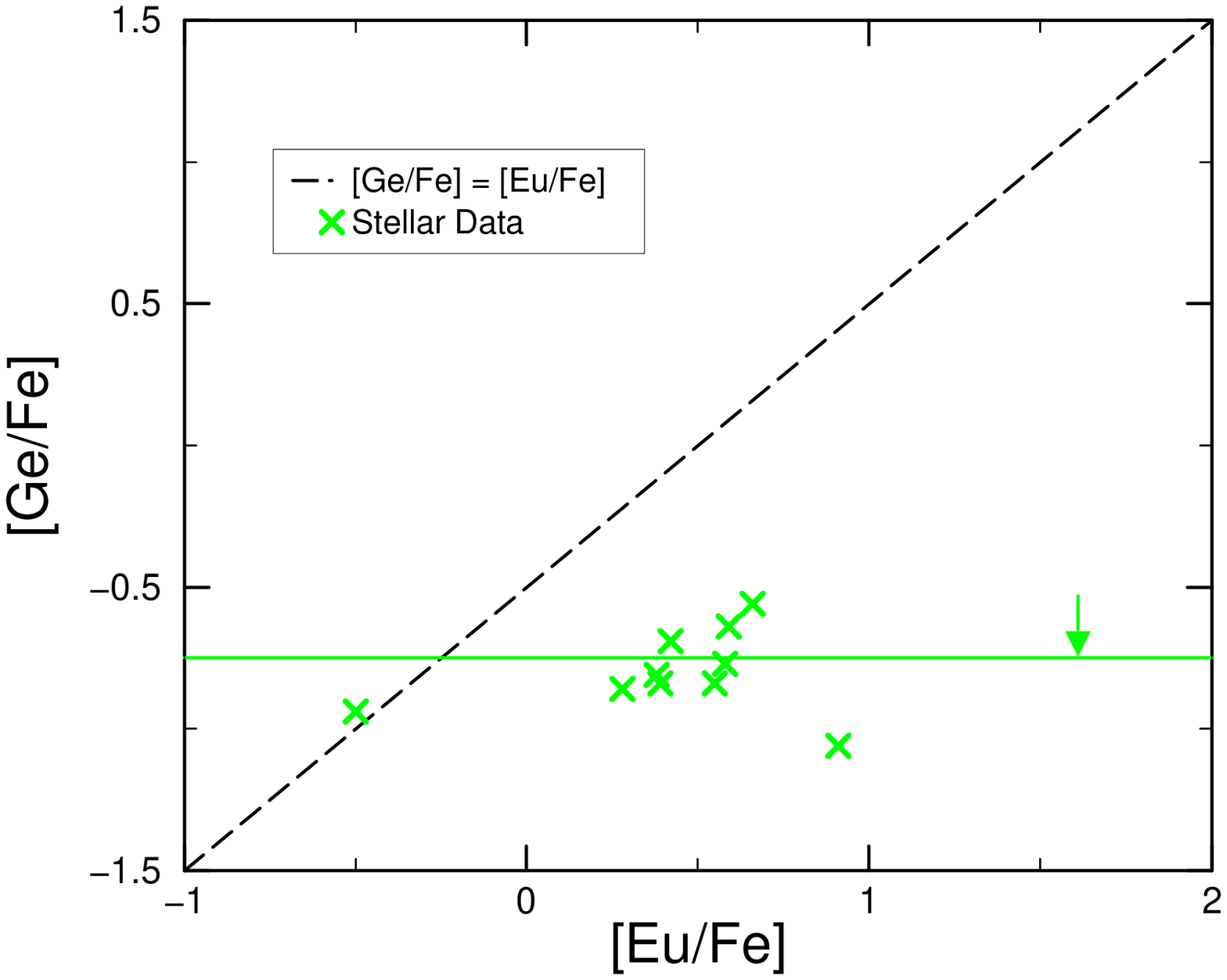}
\caption{(left) Relative abundances [Ge/H] displayed as a function of 
metallicity for our sample of 11 Galactic halo stars.
The arrow represents the derived upper limit for \cs22.
The dashed  line indicates the solar abundance ratio of these elements,
[Ge/H]~=~[Fe/H], while the solid  line shows the derived correlation
[Ge/H]~= [Fe/H]~--~0.79. 
(right)   
Correlation between the abundance ratios [Ge/Fe] and [Eu/Fe]
-- obtained with ground-based observations
by Simmerer et al. (2004).
The dashed line indicates a direct correlation
between Ge and Eu abundances.
As in the previous figure, the arrow represents the derived upper
limit for \cs22.
The solid  line at [Ge/Fe]r~--~0.79 is a fit to the observed data.
}
\label{fig1}
\end{figure}

In Figure \ref{fig1} we plot the [Ge/H] values (obtained with HST) 
with respect to
metallicity,  [Fe/H], for the 11 Galactic halo stars. 
The figure makes clear that there is a direct correlation between Ge and
Fe but at a depressed level, [Ge/H] = [Fe/H] -- 0.79,
with respect to the solar value.
This element,  normally thought of as an $n$-capture element,  appears to
have been synthesized in a different manner for low
metallicities, early in the history of the Galaxy. Comparison of the
Ge and Eu abundances, obtained with ground-based observations 
(Simmerer et al. 2004),  shows no correlation in these stars
(see Figure \ref{fig1} and Cowan et al.
2004).
The abundance data might be explained  if
Ge were synthesized in charged-particle reactions,
for example,
during the so-called ``$\alpha$-rich freeze-out'' in a supernova
environment.
Since solar system material shows clear signatures of $n$-capture for this
element (Simmerer et al. 2004), 
it is expected that at higher metallicities 
than those studied here,
and with the onset 
of the $s$-process in the Galaxy,  
Ge production would no longer  correlate with iron abundance.

\begin{figure}
\epsfxsize=5.2in
\plottwo{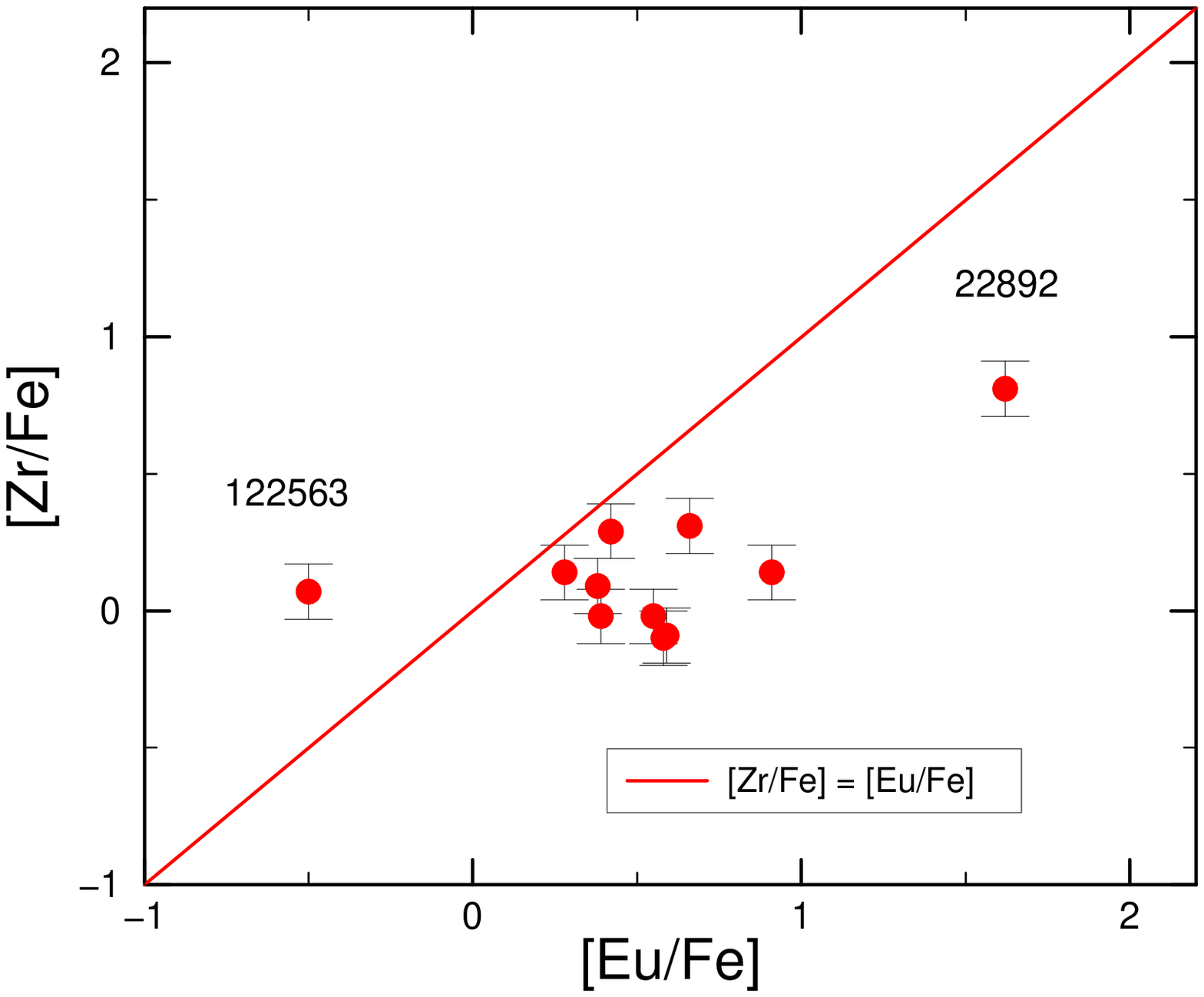}{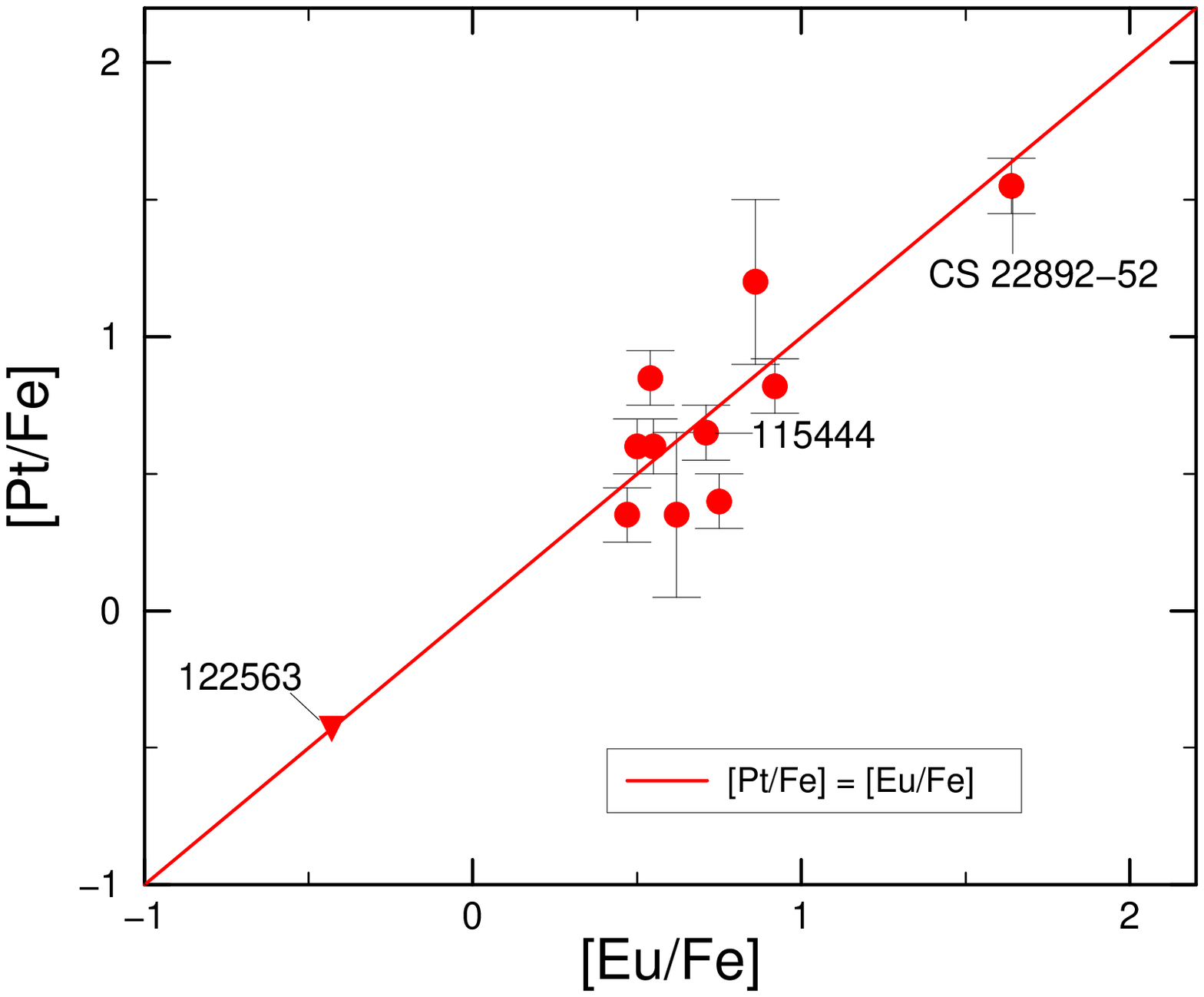}
\caption{ 
The ratio of [Zr/Fe] (left) and 
[Pt/Fe] (right), obtained with HST,
is compared  to   [Eu/Fe] 
in 11 Galactic halo stars.   
The abundance data do not show a correlation between Zr and Eu, but
clearly do for Pt and Eu.
\label{fig2}}
\end{figure}

Similar abundance comparisons with Zr (see Figure \ref{fig2})
show little  correlation with either iron
or with Eu. 
The one exception is \cs22, for which [Zr/Fe] is much  higher than
in, for example, HD~122563. \cs22 also has one of the largest 
[Eu/Fe] abundances found in halo stars.   
Our abundance determinations in these target 
stars, as well as more extensive abundance analyses 
by Travaglio et al. (2004),  suggest $n$-capture processes are responsible for 
some of the production of Zr, but that the
nucleosynthetic origin of this element is different than that for heavier
$n$-capture elements such as Ba and Eu. There are even some indications
that some type of a (lighter element)
primary process is also responsible for some fraction of the synthesis of
this element, 
as well as Sr and Y.

We also show  in Figure \ref{fig2} the ratios  of 
[Pt/Fe] as a function of
[Eu/Fe] in our sample stars. 
It is very clear in the figure that there is a direct correlation
between the Pt and Eu abundances - indicating a similar origin for
both of these $n$-capture elements.  
Os and Ir  abundance detections also show a similar
pattern with metallicity
and demonstrate a correlation between  Os, Ir, Pt  and Eu.
Our very-heavy-element abundance comparisons strongly suggest a similar
synthesis origin for Eu, Os, Ir, and Pt in the $r$-process sites that were
the progenitors of  the observed halo stars.

\section{Galactic Abundance Scatter}

Additional abundance comparisons indicate striking differences in the 
elemental abundance scatter among the elements Ge, Zr, Os, Ir and Pt.
We find essentially no scatter in [Ge/H] as a function of [Fe/H]
over the metallicity range studied (--3.1 $<$ [Fe/] $<$ -1.5). 
The Zr abundances show little scatter with the exception of \cs22.
It has been shown previously, however, that [Eu/Fe] shows large 
scatter, particularly at very low metallicities, early in the history of
the Galaxy. Our new \third \ r-process abundance determinations 
indicate,  for the {\it first} time,
 that Os-Ir-Pt show a similar abundance scatter,
tracking the [Eu/Fe] abundances in these halo stars (Cowan et al. 2004). 
This is an additional indication that Eu, Os, Ir and Pt have a similar
nucleosynthesis origin. 

\section{Elemental Abundance Distribution in \cs22}

\begin{figure}[!ht]
%\plotfiddle{22sol10.ps}{2.5in}{0}{0.5}{0.5}{0pt}{0pt}
%\epsfxsize=2.5in
\epsfysize=3.0in
\plotone{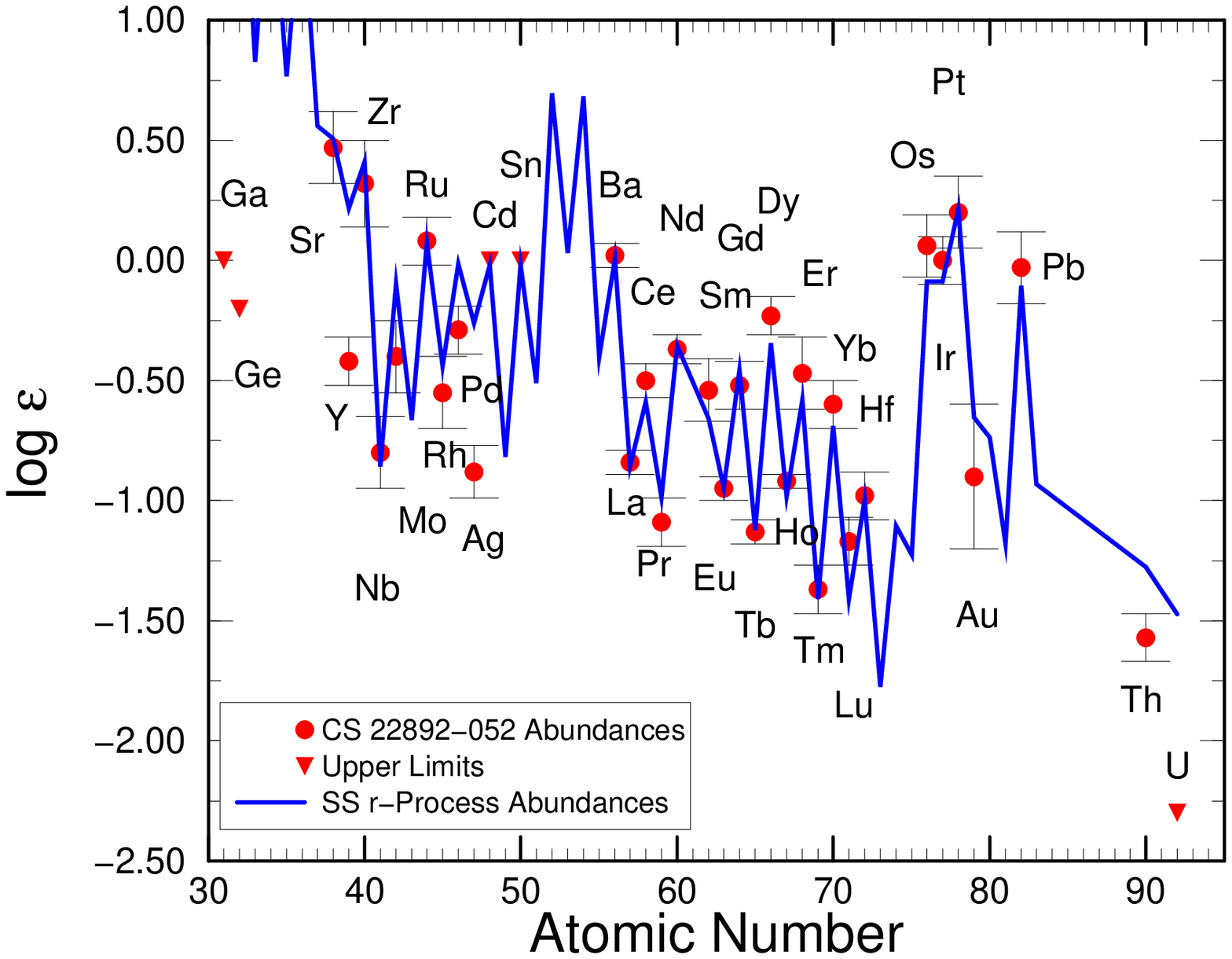}
\caption{
The neutron-capture elemental abundance
pattern
in the Galactic halo stars CS 22892--052 
compared with 
the (scaled) solar system {\it r}-process abundances (solid line).
}
\label{fig3}
\end{figure}

One of the stars in our survey is the well-studied, very metal-poor
([Fe/H] = --3.1) Galactic halo giant \cs22.
Our HST and Keck 
observations have detected the \third \ $r$-process peak elements 
Os, Ir and Pt in \cs22.  
The abundances of these elements fall on the same scaled solar system
$r$-process curve that also matches the abundances of the rare-earth 
elements (Sneden et al. 2003) such as Eu,    
as shown in Figure~\ref{fig3}. This re-emphasizes  the  common origin for all 
four elements.  
The detailed elemental abundance 
distribution for this star  
employs the new atomic experimental data for Pt (Den Hartog \etal\ 2004) and 
more reliable determinations for 
Nd (Den Hartog \etal\ 2003) and Ho (Lawler, Sneden,
\& Cowan 2004).
The solar system curve (solid line in Figure~\ref{fig3}) 
was   
determined based upon the classical $s$-process model and
utilizing the most recent $r$-/$s$-process deconvolution 
(Simmerer \etal\ 2004). 
\setlength{\textwidth}{3.5in}
\setlength{\textwidth}{5.25in}

While it is clear from the figure that abundances of the elements from Ba (Z=56)
through the \third \ $r$-process peak are all consistent with the solar
$r$-process abundances, the lighter $n$-capture elements -- including upper 
limits for Ga and Ge --   in general fall below 
that same solar curve. This might indicate two different sets of conditions,
or perhaps separate sites, for the synthesis of the heavier and lighter
$n$-capture elements (see further discussion in Sneden \& Cowan 2003). 
We note that the new HST Zr abundances are consistent with previous
ground-based determinations (Sneden et al. 2003) for this element. 

\begin{quote}
{\bfseries Acknowledgments.} We thank our colleagues for their help and
advice. 
This work has been supported in part by NSF grants
AST 03-07279 (J.J.C.)  and AST 03-07495 (C.S.),
and by STScI grants GO-8111 and GO-8342.

\end{quote}


\begin{thebibliography}{}

\bibitem[]{cow04a}
Cowan, J. J., et al.  2004, ApJ, submitted 

\bibitem[]{cow04b}
Cowan, J. J., \& Thielemann, F.-K., 2004, Phys. Today, in press


\bibitem{den03}
Den Hartog, E. A., Lawler, J. E., Sneden, C., \& Cowan, J. J. 2003,
\apjs,  148, 543

\bibitem{den04}
Den Hartog, E. A.,  Herd, T. M., Lawler, J. E., Sneden, C.,
Cowan, J. J., \& Beers, T. C. 2004, \apj, in press 

\bibitem{law04}
Lawler, J. E., Sneden, C., \& Cowan, J. J. 2004, \apj, 608, 850

\bibitem{sim04}
Simmerer, J., Sneden, C.,  Cowan, J. J., Collier, J.,  Woolf, V.,
\& Lawler, J. E. 2004, \apj, in press

\bibitem{sne03a}
Sneden, C., \& Cowan, J. J. 2003, Science, 299, 70

\bibitem{sne03}
Sneden, C.,  \etal\ 2003, \apj, 591, 936

\bibitem{tra04}
Travaglio, C.,  Gallino, R.,  Arnone, E., Cowan, J. J., Jordan, F.,
\& Sneden, C.  2004, \apj,  601, 864

\bibitem{tru02}
Truran, J. W., Cowan, J. J., Pilachowski, C. A., \& Sneden, C.
2002, \pasp,  114, 1293

\end{thebibliography}
\end{document}